\documentclass{aa}  
\usepackage{graphicx}
\usepackage{multirow}
\usepackage[colorlinks=true,linkcolor=blue]{hyperref}
\usepackage{threeparttable}
\usepackage{txfonts}
\begin{document}

            \title{Impacts of the image alignment over frequency for VLBI Global Observing System}

  \author{Ming H. Xu
          \inst{1,2,3}
          \and
          Tuomas Savolainen\inst{1,2,4}
          \and
          James M. Anderson\inst{3,6}
          \and
          Niko Kareinen\inst{5}
          \and
          Nataliya Zubko\inst{5}          
          \and 
          Susanne Lunz\inst{6}
          \and
          Harald Schuh\inst{6,3}
          }

   \institute{Aalto University Mets\"{a}hovi Radio Observatory, Mets\"{a}hovintie 114, FI-02540 Kylm\"{a}l\"{a}, Finland\\
\email{minghui.xu@aalto.fi}
\and
Aalto University Department of Electronics and Nanoengineering, PL15500, FI-00076 Aalto, Finland
\and
Technische Universit\"{a}t Berlin, Institut f\"{u}r Geod\"{a}sie und Geoinformationstechnik, Fakult\"{a}t VI, Sekr. KAI 2-2, Kaiserin-Augusta-Allee 104-106, DE-10553 Berlin, Germany
\and
Max-Planck-Institut f\"{u}r Radioastronomie, Auf dem H\"{u}gel 69, DE-53121 Bonn, Germany
\and
Finnish Geospatial Research Institute, Geodeetinrinne 2, FI-02430 Masala, Finland
\and
Deutsches GeoForschungsZentrum (GFZ), Potsdam, Telegrafenberg, DE-14473 Potsdam, Germany
}
   \date{Received ***; accepted ***}

\titlerunning{Source positions determined by VGOS observations}

 
  \abstract
 {}
   {The VLBI Global Observing System, which is the next generation of geodetic VLBI and is called VGOS, observes simultaneously at four frequency bands in the range 3.0--10.7\,GHz (expected to be extended to 14 GHz). Because source structure changes with frequency, we aim to study the source position estimates from the observations of this new VLBI system.}
   {Based on an ideal point source model, simulations are made to determine the relation between the source positions as determined by VGOS observations and the locations of the radio emission at the four bands.}
   {We obtained the source positions as determined by VGOS observations as a function of the source positions at the four frequency bands for both group and phase delays. The results reveal that if the location of the radio emission at one band is offset with respect to that at the other bands, the position estimate can be shifted to the opposite direction and even by more than three times that offset.}
   {The VGOS source positions will be very variable with time and very imprecise in the sense of relating to the locations of the radio emission at the four bands, if the effects of source structure are not modeled. The image alignment over frequency is essential in order to model the effects of source structure in VGOS observations, which is the only way to mitigate these strong frequency-dependent impacts on VGOS source positions.}

   \keywords{reference systems / astrometry / galaxies: active / galaxies: jets / radio continuum: galaxies}

   \maketitle
%

\section{Introduction}

In fundamental astronomy, two cardinal improvements in the precision of astrometric measurements of positions of celestial objects have been achieved based on very long baseline interferometry (VLBI) at radio wavelengths and the European Space Agency mission Gaia\footnote{\url{https://sci.esa.int/web/gaia}} \citep{2016A&A...595A...2G} at optical wavelengths. A detailed comparison between the third realization of the International Celestial Reference Frame (ICRF) \citep[ICRF3;][]{ICRF3_2020} and the Gaia Early Data Release 3 \citep[EDR3;][]{2020arXiv201201533G} shows that the median difference of radio and optical source positions for more than 2000 common sources is about 0.5\,milliarcsecond (mas) and for the sources with optical $G$ magnitudes < 18.0 mag the median difference is on the order of 0.3\,mas \citep{2021A&A...647A.189X}. On the other hand, for the sources that have radio to optical position differences larger than 3$\sigma$, which is $\sim$24\% of the common sources, it has become more and more convincing that the position differences are parallel to the directions of the radio jets \citep{2017A&A...598L...1K,2017MNRAS.467L..71P,2019ApJ...871..143P,2021A&A...647A.189X}. Our recent work showed that these sources are more likely to have extended structure at radio wavelengths \citep{2021A&A...647A.189X}. VLBI and Gaia already have the potential of detecting astrophysical properties about the radio and optical emission at mas scales for these objects, which are mostly active galactic nuclei (AGNs) \citep{2019ApJ...871..143P,2019MNRAS.482.3023P}.  

In the last several decades, celestial reference frame (CRF) sources were observed predominantly at S/X bands by geodetic VLBI to derive their positions for building the ICRFs \citep{1998AJ....116..516M,fey2015aj,ICRF3_2020}. The geodetic VLBI observations are coordinated by the International VLBI Service for Geodesy and Astrometry \citep[IVS\footnote{\url{https://ivscc.gsfc.nasa.gov/index.html}};][]{schuh2012jg,nothnagel2017jg} since 2000. In order to achieve 1~mm station position accuracy and 0.1~mm/yr velocity stability on global scales, 
the IVS is developing the next-generation, broadband geodetic VLBI system, known as the
VLBI Global Observing System \citep[VGOS;][]{niell2007,2009vlbi.rept....1P}. Even though the primary goals are to greatly increase the precision of Earth orientation parameters and the International Terrestrial Reference Frame (ITRF) \citep{altamimi2016jgr} by using VGOS, the CRF as an integral part of the geodetic VLBI is also expected to be --- and needs to be --- improved in the VGOS era. Based on 21 actual VGOS sessions, it is demonstrated that the measurement noise in VGOS group delay observables is at the level of $\sim$2\,ps \citep{2021arXiv210212750X}. A network of nine VGOS stations has observed a 24-hour session biweekly in 2019 and 2020 and will start to observe weekly in the near future; this network is expanding globally as planned \citep{2020JGeod..94..100B}, with another eleven stations built and nine stations in the planning stage as of January 2021. VGOS observations started to contribute to the building of the ITRF through the geodetic solutions submitted by the IVS analysis centers in 2020\footnote{\url{https://ivscc.gsfc.nasa.gov/IVS_AC/IVS-AC_ITRF2020.htm}}. 

Currently, the VGOS system observes simultaneously at four 512 MHz wide bands centered at 3.3, 5.5, 6.6, and 10.5 GHz, which are labeled as band A, B, C, and D, respectively, with 32 recording channels (see \citet{niell2018rs} and Table \ref{10GHz} for the channel frequencies used in the current VGOS observations), while the legacy S/X system observes at S ($\sim$2.2\,GHz) and X ($\sim$8.6\,GHz) bands. There is a substantial difference in the observing frequencies and thus the received radio signals from AGNs between the legacy system and the VGOS system.
The CRF sources have different structure at different bands and the structure changes over time at angular scales of sub-mas, therefore, we cannot assume that the locations of the radio emission of a source at different frequency bands are at the same position or have stable relative positions between the four bands at angular scales of $\sim$0.1\,mas to mas. The fact that source positions change with frequency is one of the reasons for which the ICRF3 has three separate catalogs at the three frequency bands \citep{ICRF3_2020}. The fundamental question is where the position of a source determined by VGOS observations with respect to the locations of the radio emission at the four bands is. We aim to address this question in the study.

On the other hand, because the impact of source structure on VGOS group delay observables is about one order of magnitude larger than the random noise level as shown in \citet{2021arXiv210212750X}, the IVS is making effort, towards deriving source-structure corrections for geodetic solutions by collecting information of antenna system temperatures and gain curves and making images from VGOS observations. Before VGOS source-structure corrections are generated, it is necessary that the images at the four bands are aligned with respect to each other. The impact of the potential errors in that alignment needs to be studied. This is equivalent to the question that we aim to discuss.

The purposes of this study are: (1) to investigate the potential impacts of the effects of source structure on the source positions as determined by VGOS observations if these effects will not be modeled and (2) to demonstrate the importance of the image alignment over frequency when one wants to correct for these effects. The paper is structured as follows. We introduce in Sect. 2 the changes of source positions with respect to frequency based on the ICRF3 and the images obtained based on actual VGOS images. In Sect. 3 we first describe how the variations in the channel phases affect the VGOS observables and then derive the formula of the position estimate from VGOS observations as a function of the locations of the radio emission at the four bands. We make the discussion in Sect. 4 and our conclusion in Sect. 5.
\section{Changes of source positions over frequency}\label{sect2}
Based on globally absolute astrometric observations by VLBI at S/X, K, and X/Ka bands, the ICRF3 was established independently at these three frequencies \citep{ICRF3_2020}. The two catalogs at K and X/Ka bands were aligned to the S/X catalog by applying no-net-rotation constraints, leading to non-significant spin-rotations between them. However, the glide deformation between the S/X and K catalogs is 10--30 microarcseconds ($\mu$as), and that between the S/X and X/Ka catalogs is as large as $\sim$300 $\mu$as. These deformations were believed to be due mainly to the different observing networks among the three bands, e.g., the X/Ka observations were made from a network of four stations, instead of the intrinsic source position differences. After applying transformations including both the first-degree and the second-degree deformation parameters (in total 16 parameters) fitted from the position differences between these three catalogs, the median angular separation of the K-band positions relative to the S/X-band positions is $\sim$0.2~mas, the same level as that for the X/Ka-band positions; about 6\% of the common sources in the K and S/X catalogs have position differences significant at the 3$\sigma$ confidence level, and 11\% for the common sources in the X/Ka and S/X catalogs.

\begin{figure*}[tbhp!]
   \centering
            \includegraphics[width=0.47\textwidth]{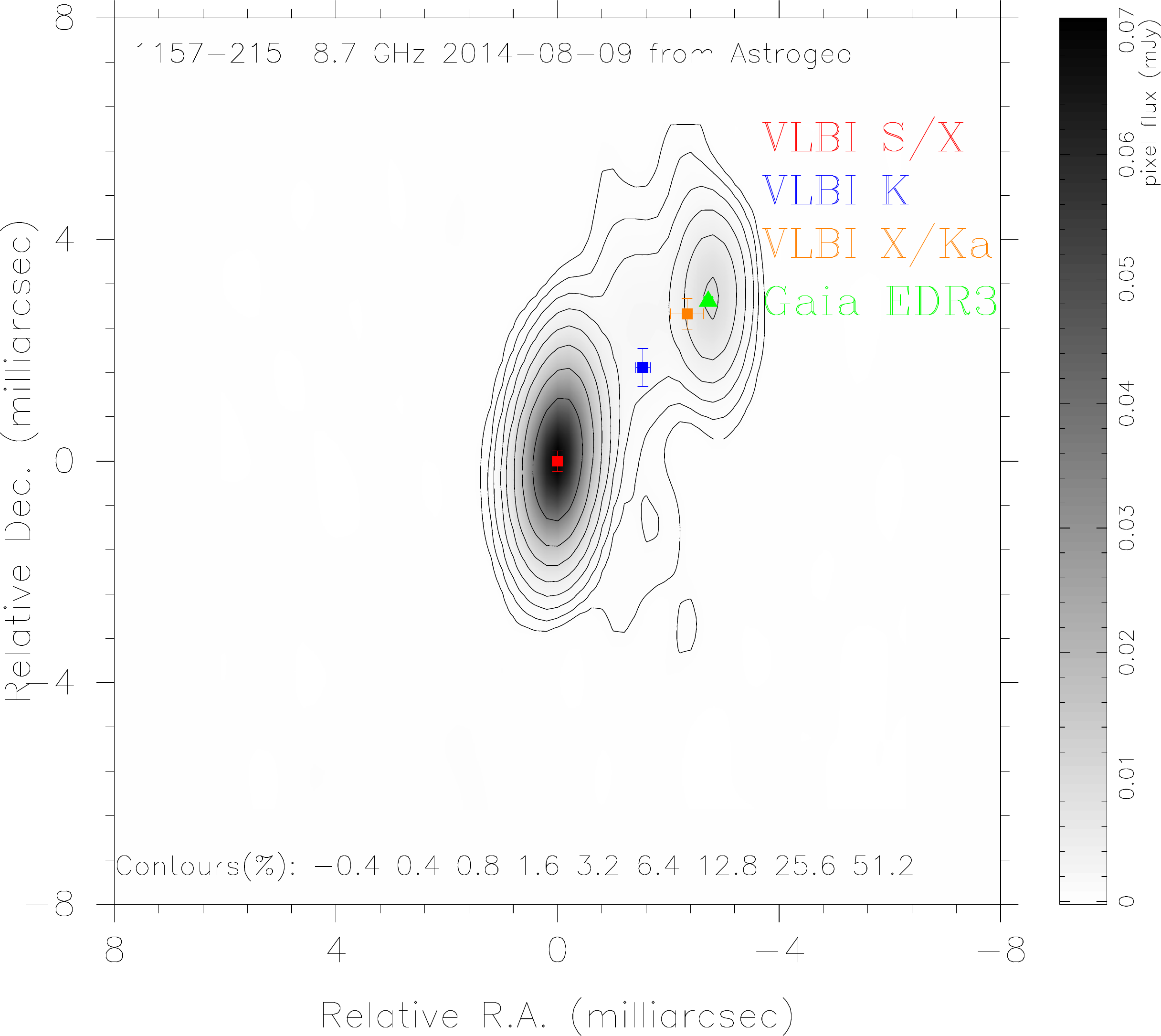}
         \includegraphics[width=0.47\textwidth]{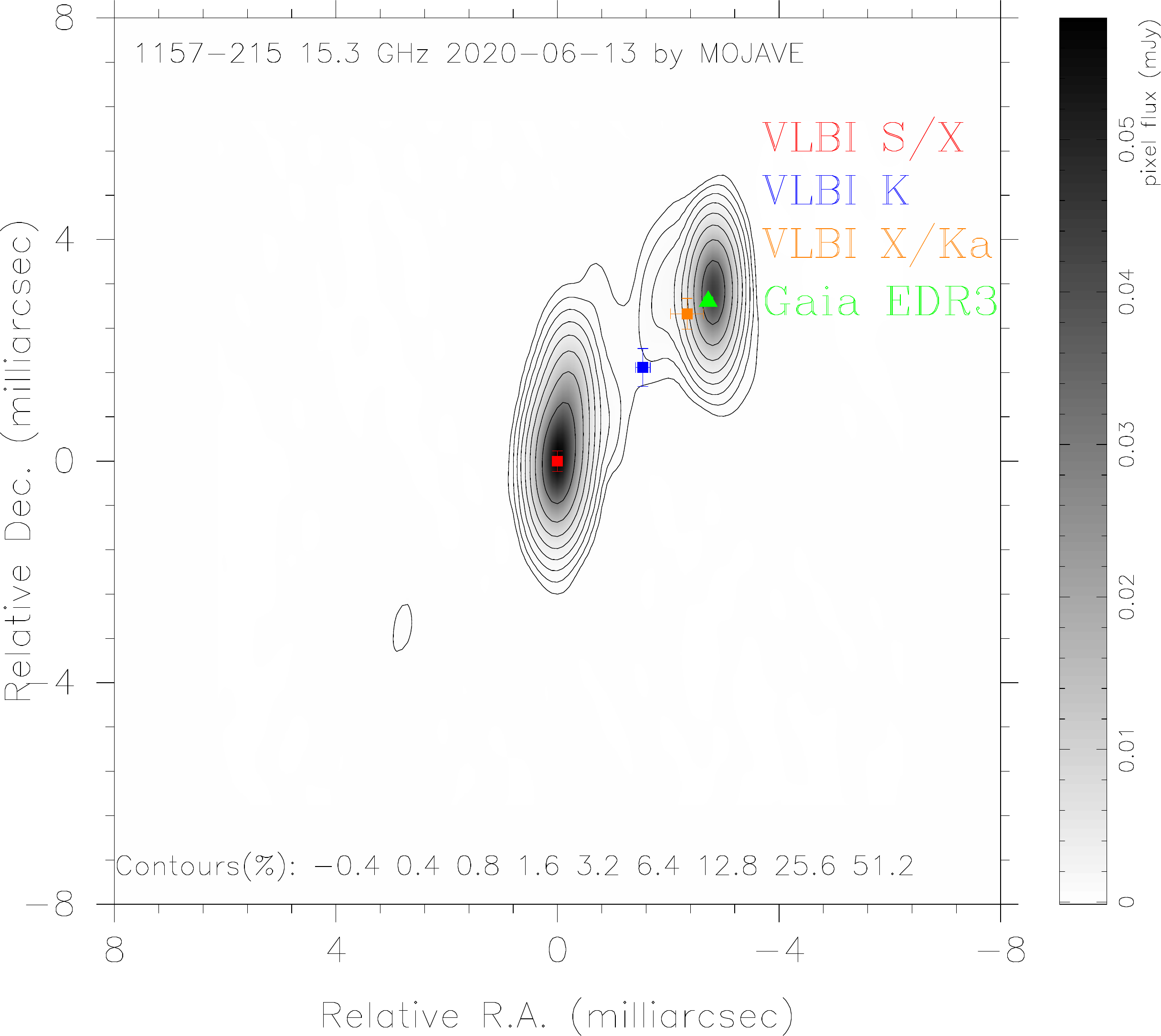}
   \caption{Three radio positions from the ICRF3 at the S/X, K, and X/Ka bands and the Gaia position for source 1157$-$215, on its radio images at 8.7\,GHz from Astrogeo (left) and at 15.3\,GHz from MOJAVE (right). Overlay contours are shown at levels of peak percentage specified in the bottom of plot. Since these radio images do not have information of the absolute positions, the peaks are formally assumed to be located at the S/X-band positions. Error bars shown are the uncertainties from the four catalogs. The K- and X/Ka-band positions are the ones after the full deformation transformations \citep{ICRF3_2020}. The angular separations of the K- and X/Ka-band positions and the Gaia position relative to the S/X-band position for this source are 2.29$\pm$0.31\,mas, 3.54$\pm$0.32\,mas, and 3.97$\pm$0.13\,mas, respectively.}
              \label{src_1157-215}%
    \end{figure*}

Figure \ref{src_1157-215} shows another example for the source 1157$-$215 with two components. The ratio of the flux densities between the southeastern component and the northwestern component largely increases from 15.3\,GHz to 8.7\,GHz; this shifts the radio positions towards the southeast direction when frequency decreases. The northwestern component is more likely to be the core than the brightest component in these images. These two sources have angular separations of the positions in the K, X/Ka, and Gaia catalogs relative to that in the S/X catalog all significant at the 3$\sigma$ confidence level (see Tables 14, 15, 16 in \cite{ICRF3_2020}). Figure \ref{src_1157-215} demonstrates that source structure and source positions change significantly with frequency and for this source the radio positions move in the directions of the optical positions when the radio frequency goes higher. The changes of the peak component among the core and the jet components at various frequency bands will lead to large position offsets between different frequencies, as the source position referred to by VLBI observations in general is dominated by the position of the peak. 

%

\begin{figure*}[tbhp!]
   \centering
            \includegraphics[width=0.45\textwidth]{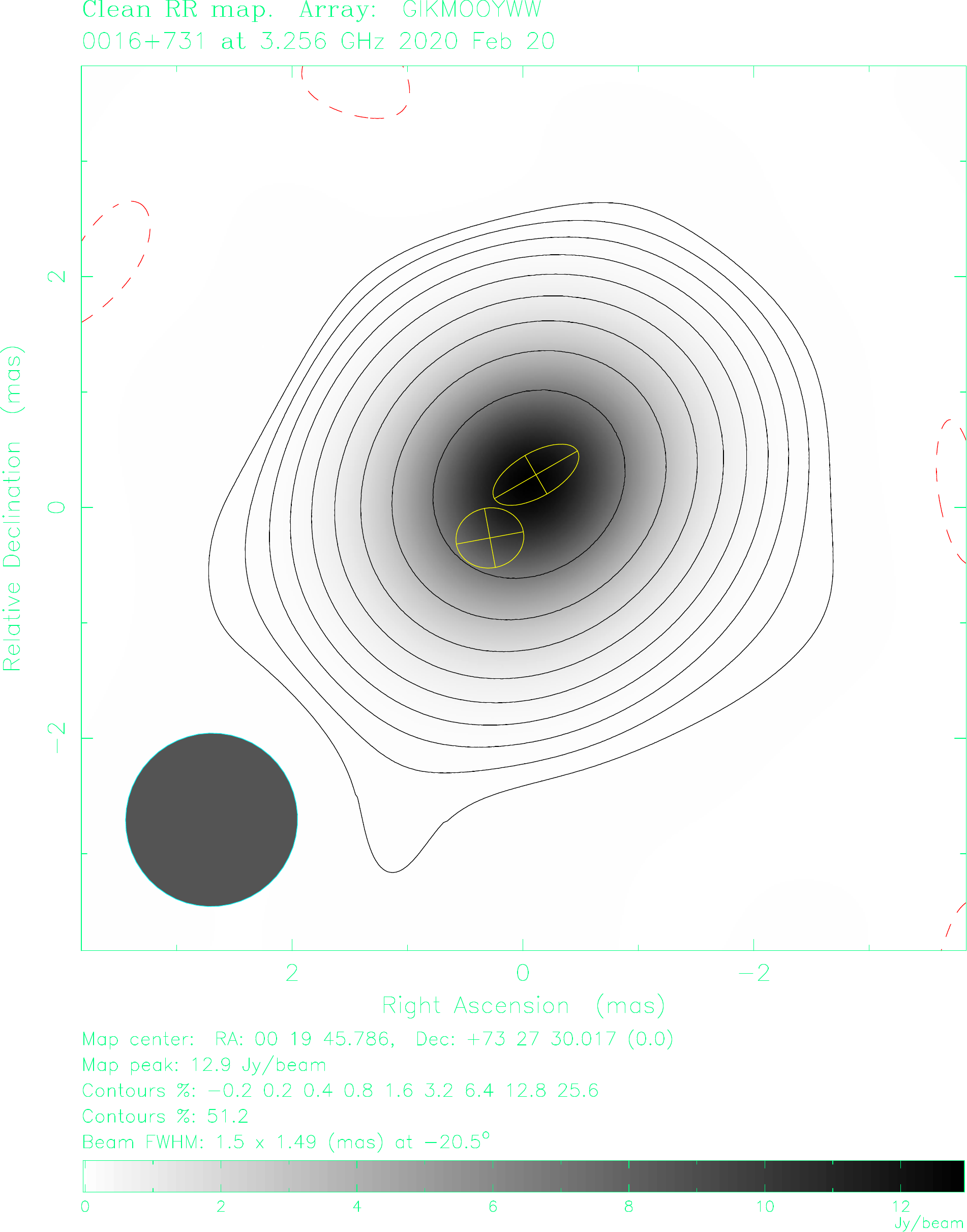}
         \includegraphics[width=0.45\textwidth]{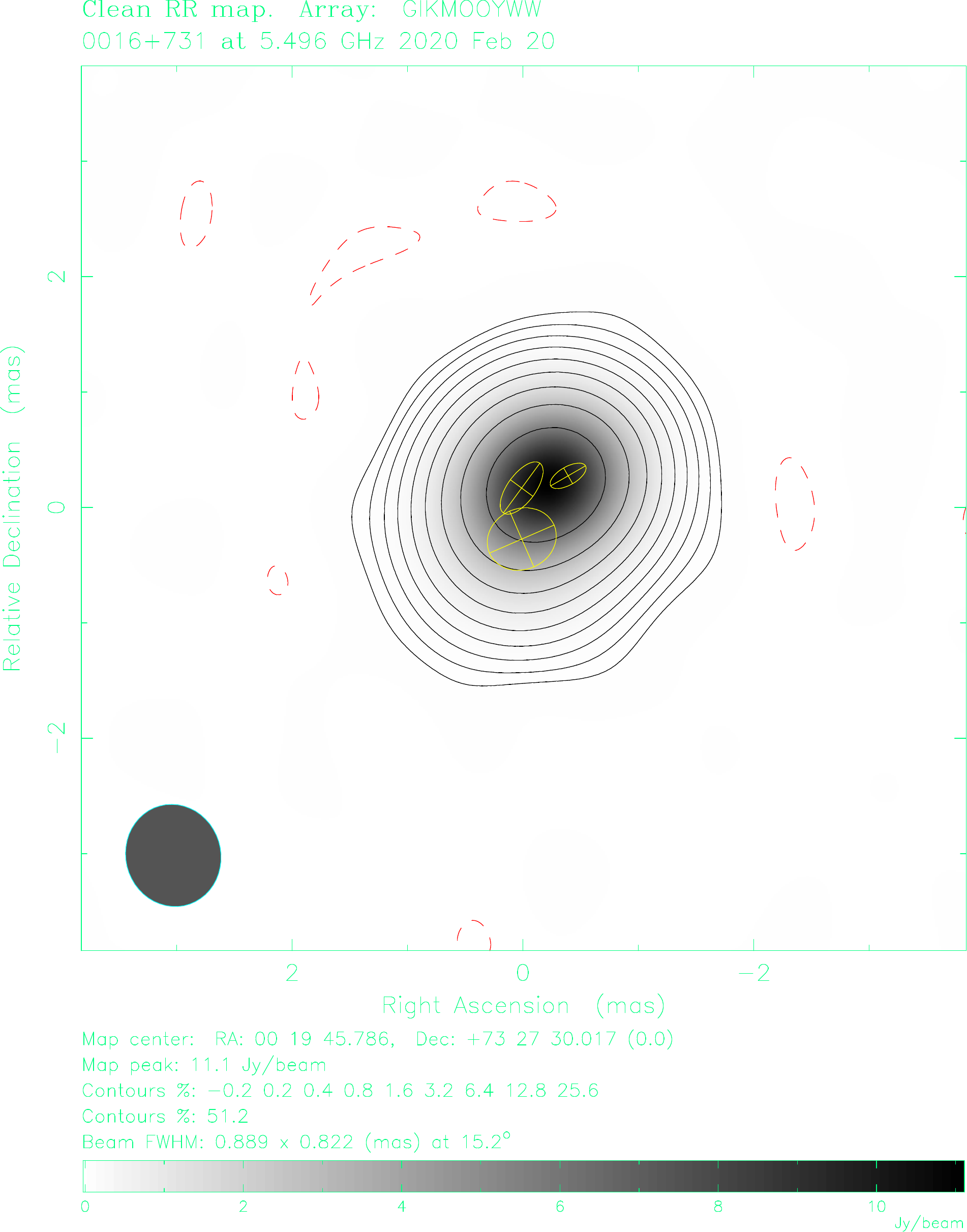}
                     \includegraphics[width=0.45\textwidth]{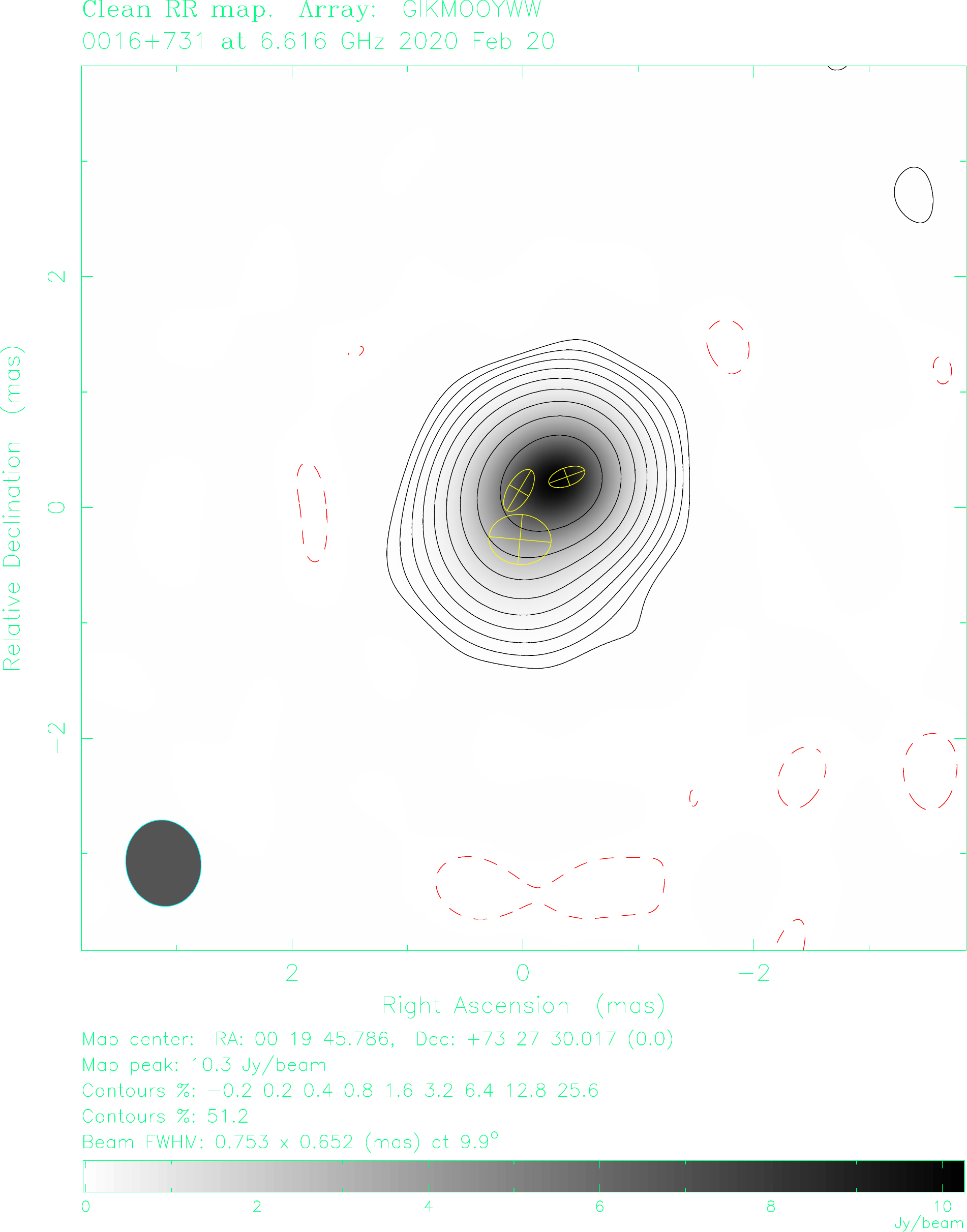}
         \includegraphics[width=0.45\textwidth]{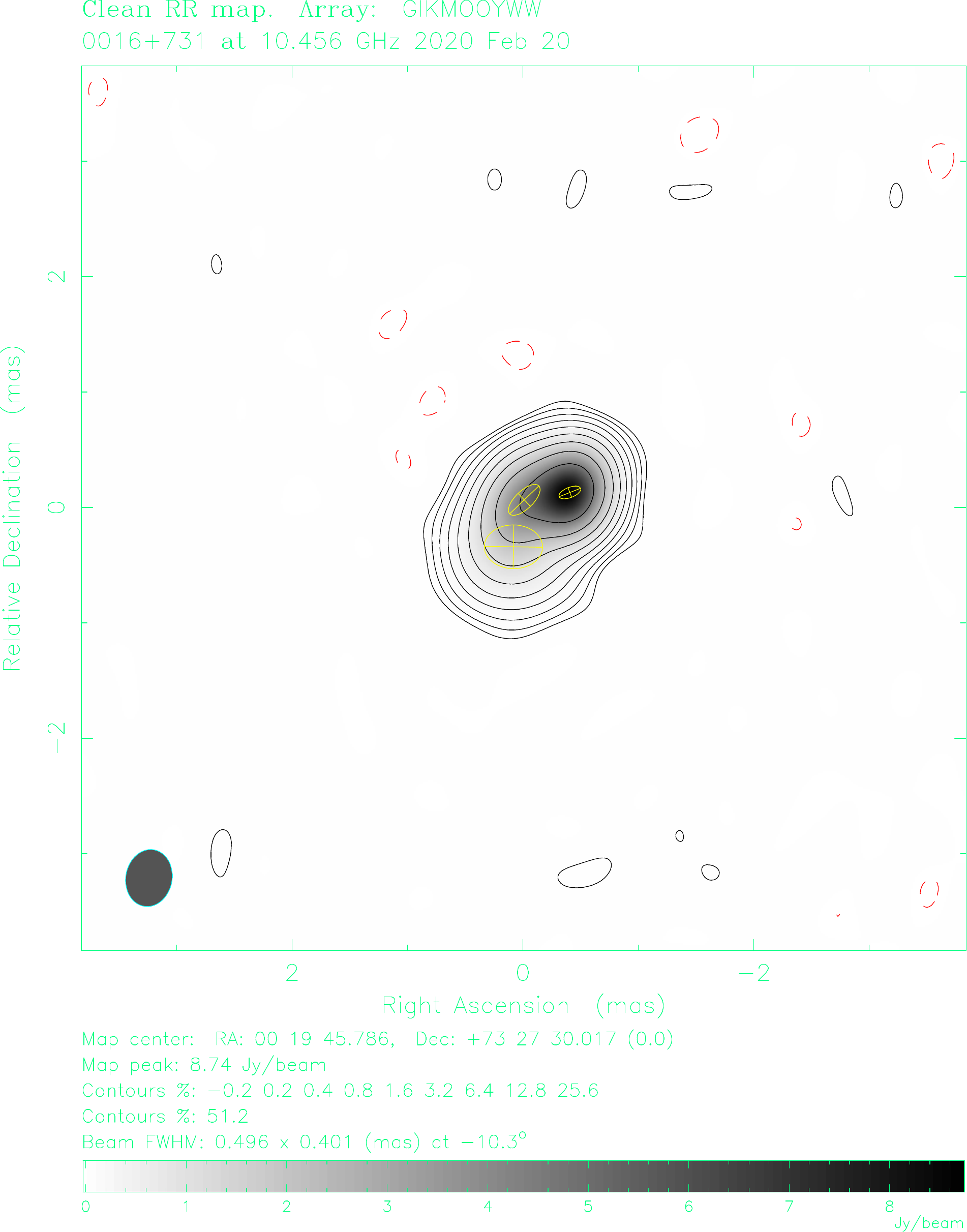}
   \caption{Images of source 0016+731 from VGOS session VO0051 (Feb. 20, 2020) at the frequencies 3.3 GHz (upper left), 5.5 GHz (upper right), 6.6 GHz (bottom left), and 10.5 GHz (bottom right). The yellow ellipses indicate the Gaussian components modeled by difmap. These images are constructed from the VGOS observations, which are calibrated based on the images derived from closure phases and closure amplitudes \citep{https://doi.org/10.1029/2020JB021238}. Therefore, the units of the pixel flux densities are arbitrary. The beam size is shown as a black ellipse in the bottom left corner of each plot.}
              \label{Image_0016+731}%
    \end{figure*}

In general cases, however, the cores of geodetic radio sources can be reliably identified because one can rely on a database with the images of a long time series and the spectrum index maps. Moreover, these sources tend to have compact cores with extended but weaker jets. To investigate these general cases, we use the images constructed from actual VGOS observations by:
   \begin{enumerate}
\item deriving the closure images by using the method in \citet{https://doi.org/10.1029/2020JB021238};
\item calibrating VGOS observations based on the closure images;
\item performing model fitting by using difmap.
\end{enumerate}
Figure \ref{Image_0016+731} shows the images of source 0016+731 from model fitting in difmap at the four frequency bands based on the VGOS observations VO0051. The source 0016+731 has a compact core, the northwestern component, and weaker jet components at all the four bands. Three components are consistently detected at the highest three frequencies, whereas only two components can be detected at the lowest frequency band due to much lower angular resolution. At 10.5\,GHz, the angular distance between the core and the closest jet component is about 0.44\,mas with a flux density ratio of 0.53, and the angular distance from the second jet component to the core is about 0.68\,mas with a flux density ratio of 0.25. At 3.3\,GHz, the angular distance between the core and the jet component is about 0.68\,mas with a flux density ratio of 0.20. With an increase in the angular resolution by a factor of twofold to threefold at the three highest frequency bands, the core at 3.3 GHz is resolved. Based on the parameters of the Gaussian components at 10.5\,GHz, the position of the core would be shifted towards east by 0.14\,mas due to the contribution of the nearest jet component if it appeared as the same structure at 3.3\,GHz and 10.5\,GHz. Furthermore, if the effects of source structure are not modeled, the source position shifts due to the contribution of the jet components will happen at all the frequency bands with larger magnitudes. It is obvious that there are source position offsets over frequency, since the angular resolutions of simultaneous observations at the multiple frequency bands are significantly different, leading to different contributions of the jet components to the cores at various frequency bands.

Changes of source positions with frequency are due mainly to two factors: extended source structure and the frequency-dependence of the core position, the so called core shift~\citep{1979ApJ...232...34B}. This is supported by the ICRF3 multi-frequnecy catalogs and the optical positions from Gaia. It is well demonstrated that the differences between the radio positions and the optical positions are parallel to the jet directions \cite[see, e.g.,][]{2021A&A...651A..64L,2021A&A...652A..87L}. These two effects in general are larger at the lower frequency bands (the VGOS frequency bands) than at the higher frequency bands (K and X/Ka bands). It is expected, however, that the impact of these effects at S band on the previous ICRFs is significantly reduced. According to the linear combination of the delay observables at S and X bands to remove the ionospheric effects, which is done in geodetic data analysis after the fringe fitting process, the source position determined from VLBI observations at these two bands, denoted by $\vec{k}_{\scriptsize \mbox{S/X}}$, is given approximately by
\begin{equation}
\label{SX_position}
\vec{k}_{\scriptsize \mbox{S/X}}= 1.07\vec{k}_{\scriptsize \mbox{X}}-0.07\vec{k}_{\scriptsize \mbox{S}},
\end{equation}
where $\vec{k}_{\scriptsize \mbox{X}}$ and $\vec{k}_{\scriptsize \mbox{S}}$ are the group-delay source positions at the two bands \cite[see, e.g.,][]{porcas2009}. When a source is ideal point-like at S and X band, these two positions can be located at the same direction of the jet base \citep{porcas2009}; in general cases, they are in different directions. As shown in Eq. \ref{SX_position}, the contribution of the S-band positions and their variations are reduced by a factor of 14. We should note that the effects of source structure at S band are also scaled down by that factor. In the legacy VLBI system, the source positions at the higher band dominate over that at the lower band. In the VGOS system, however, observations are made at four frequency bands instead of two bands, and the ionospheric effects are fitted in the fringe fitting process through which group delay observables are determined. As a consequence, VGOS source positions need to be studied thoroughly.
 
\section{Simulation for VGOS source positions}\label{sect3}
\subsection{Changes in broadband observables due to phase variations}
Let us recall the model of using 32 channel phases to determine the broadband observables as done in the VGOS post processing using \emph{fourfit}\footnote{\url{https://www.haystack.mit.edu/wp-content/uploads/2020/07/docs_hops_000_vgos-data-processing.pdf}}. Following \citet{cappallo2016}, the visibility phase $\phi_{\nu_i}$ at the channel with frequency ${\nu_i}$ can be expressed as 
    \begin{equation}
     \label{VGOS_structure_effect}
     \phi_{\nu_i}=\tau^{g} ( \nu_{i} - \nu_{0} ) + \phi_{0} - \frac{k~\delta \mbox{TEC}}{\nu_{i}},
    \end{equation}
where $\tau^{\scriptsize \mbox{g}}$, $\phi_{0}$, and $\delta \mbox{TEC}$ are the broadband group delay, the broadband phase at the reference frequency $\nu_{0}$, which is 6.0 GHz in the current data processing, and the differential total electron content (TEC) along the ray paths of the radio emission from a source to the two stations of a baseline, respectively; they are the broadband observables simultaneously fitted in \emph{fourfit}. Constant $k$ is equal to 1.3445 when phases are in units of turns of a cycle, delays in units of nanoseconds (ns), frequencies in units of GHz, and $\delta$TEC in units of TECU\footnote{1~TECU~$\equiv{10^{16}}$~electrons per square meter.}. We can derive the phase delay by $\tau^{\scriptsize \mbox{p}} = (\phi_{0}+\mbox{N})/\nu_{0}$, where $\mbox{N}$ is the integer number of phase turns and $\tau^{\scriptsize \mbox{p}}$ is in the units of ns.

We denote the observational equation as follows,
    \begin{equation}
l = Ax + \sigma,
    \end{equation}
where $l$ is the vector of the phases at the 32 channels, $A$ is the design matrix, $x$ is the vector of the three unknowns, i.e., $\tau^{g}$, $\phi_{0}$, and $\delta$TEC, and $\sigma$ is the noise vector. The vector $l$ consists of the simulated channel phases based on delay offsets or source position offsets. The design matrix $A$, with a dimension of 32 $\times$ 3, consists of the partial derivatives of phase with respect to the three unknowns, which can be derived based on Eq. \ref{VGOS_structure_effect}. By assuming that the visibility amplitudes over 32 channels are flat --- equal weights, the normal matrix can be derived by $A^{\mbox{T}}A$. By least squares fitting (LSF), the changes in the broadband observables due to changes in the channel phases can be determined. 

The results shown in Table \ref{broadband} were obtained from LSF for seven possible combinations of bands with a 1 ps delay offset based on the channel frequencies listed in Table \ref{10GHz}. The four scenarios labeled as ``Case 1'' correspond to the cases where only one of the four bands has such a delay offset causing the variations in the eight channel phases. The normal matrix in the LSF process of estimating the broadband observables is independent of channel phases, and the estimates of the broadband observables are linearly dependent on them. Therefore, there are two features in this broadband fitting process. First, the coefficients of propagating the delay offsets at individual bands to the broadband observables are independent of the magnitudes of those delay offsets: they are invariable for a given set of channel frequencies. Second, the results of all other possible combinations of bands with delay offsets can also be derived from the four basic scenarios in Case~1 through a linear combination. For instance, the first scenario in Case~2 is the summation of the results of the first two scenarios in Case~1, and the Case~3 is the summation of that of the four scenarios in Case~1. 

\begin{table*}[tbhp!]
\begin{center}
\caption{Changes in the broadband observables due to delay offsets of 1\,ps at bands of various combinations.} 
\label{broadband} \centering
\begin{tabular}{lp{0.5cm}p{0.5cm}p{0.5cm}p{0.5cm}rrrr}
\hline    \noalign{\smallskip}
 &\multicolumn{4}{c}{Delay offsets at band} && \multicolumn{3}{c}{Changes in}
\\\noalign{\smallskip}\cline{2-5}\cline{7-9}\noalign{\smallskip}
&A & B & C & D && $\tau^{\scriptsize \mbox{g}}$ & $\tau^{\scriptsize \mbox{p}}$ & $\delta \mbox{TEC}$ \\ 
& [ps]& [ps]& [ps]& [ps] &&[ps] & [ps]& [TECU]\\\hline   \noalign{\smallskip} 
\multirow{4}{*}{Case 1} &1.0 & 0.0 & 0.0 & 0.0 && $0.505$ & $-0.883$ & $-0.024$ \\
&0.0 & 1.0 & 0.0 & 0.0 && $-1.448$ & $1.729$ & $0.034$ \\
&0.0 & 0.0 & 1.0 & 0.0 && $-1.458$ & $2.044$ & $0.040$ \\
&0.0 & 0.0 & 0.0 & 1.0 && $3.401$ & $-1.889$ & $-0.050$ \\\hline  \noalign{\smallskip}
\multirow{2}{*}{Case 2} &1.0 & 1.0 & 0.0 & 0.0 && $-0.943$ & $0.846$ & $0.010$ \\
&0.0 & 0.0 & 1.0 & 1.0 && $1.943$ & $0.154$ & $-0.010$ \\\hline \noalign{\smallskip}
Case 3 &1.0 & 1.0 & 1.0 & 1.0 && $1.0$ & $1.0$ & $0.0$ \\\noalign{\smallskip}
\hline
\end{tabular}
\end{center}
\end{table*}

For general scenarios, where the delay offsets at the four bands are $\Delta\tau_{\scriptsize \mbox{A}}$, $\Delta\tau_{\scriptsize \mbox{B}}$, $\Delta\tau_{\scriptsize \mbox{C}}$, and $\Delta\tau_{\scriptsize \mbox{D}}$, respectively, the change in the broadband group delay, denoted by $\Delta\tau^{\scriptsize \mbox{g}}$, can be obtained from the results in Table \ref{broadband} as 
\begin{equation}
\label{delay_offset}
\Delta\tau^{\scriptsize \mbox{g}}=+0.505\Delta\tau_{\scriptsize \mbox{A}}-1.448\Delta\tau_{\scriptsize \mbox{B}}-1.458\Delta\tau_{\scriptsize \mbox{C}}+3.401\Delta\tau_{\scriptsize \mbox{D}},
\end{equation}
and the change in the phase delay, denoted by $\Delta\tau^{\scriptsize \mbox{p}}$, as
\begin{equation}
\label{phase_offset}
\Delta\tau^{\scriptsize \mbox{p}}=-0.883\Delta\tau_{\scriptsize \mbox{A}}+1.729\Delta\tau_{\scriptsize \mbox{B}}+2.044\Delta\tau_{\scriptsize \mbox{C}}-1.889\Delta\tau_{\scriptsize \mbox{D}}.
\end{equation}

This section is similar to what has been done for the investigation of the impact of constant instrumental delays between different bands on VGOS broadband delays by \citet{corey2018}. Equation \ref{delay_offset} can be used to recover the results of the fifteen possible scenarios in Fig. 1 and Table 2 in their VGOS memo\footnote{\url{https://www.haystack.mit.edu/wp-content/uploads/2020/07/memo_VGOS_050.pdf}}. 

\subsection{Source positions determined by VGOS observations}

The delay offset at an individual band can change with time and be baseline- and source-dependent due to some physical 
causes, for instance, a source position offset. We denote the (phase-delay) source position at band A by $\vec{k}_{0} + \Delta \vec{k}_{\scriptsize \mbox{A}}$, where $\vec{k}_{0}$ is a reference direction and $\Delta \vec{k}_{\scriptsize \mbox{A}}$ is a position offset with respect to the reference direction; and so forth $\vec{k}_{0} + \Delta \vec{k}_{\scriptsize \mbox{B}}$, $\vec{k}_{0} + \Delta \vec{k}_{\scriptsize \mbox{C}}$, and $\vec{k}_{0} + \Delta \vec{k}_{\scriptsize \mbox{D}}$ for the other three bands, respectively. These source positions are defined at the center frequencies of individual bands, because source structure and core shift evolve with frequency even within a band. We remark that in presence of source structure there is no a unique reference position of a source for different baselines or a baseline at different observing epochs, but the point here is that there are position offsets among the four bands due to extended source structure, as we have discussed in Sect. \ref{sect2}. 

The (phase) delay offset at band A due to the source position offset $\Delta \vec{k}_{\scriptsize \mbox{A}}$ is computed as 
\begin{equation}
\label{simple_delay_model}
\Delta\tau_{\scriptsize \mbox{A}}=-\frac{\vec{B}\cdot\Delta \vec{k}_{\scriptsize \mbox{A}}}{c},
\end{equation}
where $\vec{B}$ is the baseline vector and $c$ is the speed of light. There is no need need for spherical trigonometry when the position offset is small. This is always possible because we would expect the differences in the locations of the radio emission of the CRF sources between the four bands to be at the milliarcsecond level or even smaller. By applying this common 
delay/position-offset 
relation to the four delay terms in Eq. \ref{delay_offset}, the VGOS group delay position with respect to the reference direction, denoted by $\Delta\vec{k}^{\scriptsize \mbox{g}}$, is jointly determined by the locations of the radio emission at the four bands as follows:
\begin{equation}
\label{group_position}
\Delta\vec{k}^{\scriptsize \mbox{g}}= + 0.505\Delta \vec{k}_{\scriptsize \mbox{A}}-1.448\Delta \vec{k}_{\scriptsize \mbox{B}}-1.458\Delta \vec{k}_{\scriptsize \mbox{C}}+3.401\Delta \vec{k}_{\scriptsize \mbox{D}}.
\end{equation}

Similarly, the VGOS phase delay position, denoted by $\Delta\vec{k}^{\scriptsize \mbox{p}}$, is given by 
\begin{equation}
\label{phase_position}
\Delta\vec{k}^{\scriptsize \mbox{p}}=-0.883\Delta \vec{k}_{\scriptsize \mbox{A}}+1.729\Delta \vec{k}_{\scriptsize \mbox{B}}+2.044\Delta \vec{k}_{\scriptsize \mbox{C}}-1.889\Delta \vec{k}_{\scriptsize \mbox{D}}.
\end{equation}

The summation of the four coefficients in the right-hand side of Eq. \ref{group_position} is equal to unity (roundoff error of the displayed coefficients notwithstanding), as well as for Eq. \ref{phase_position}. Therefore, the position vectors $\vec{k}_{0}+\Delta\vec{k}^{\scriptsize \mbox{g}}$ and $\vec{k}_{0}+\Delta\vec{k}^{\scriptsize \mbox{p}}$ are independent of the reference direction. 

\begin{figure*}[tbhp!]
   \centering
   \includegraphics[width=0.47\textwidth]{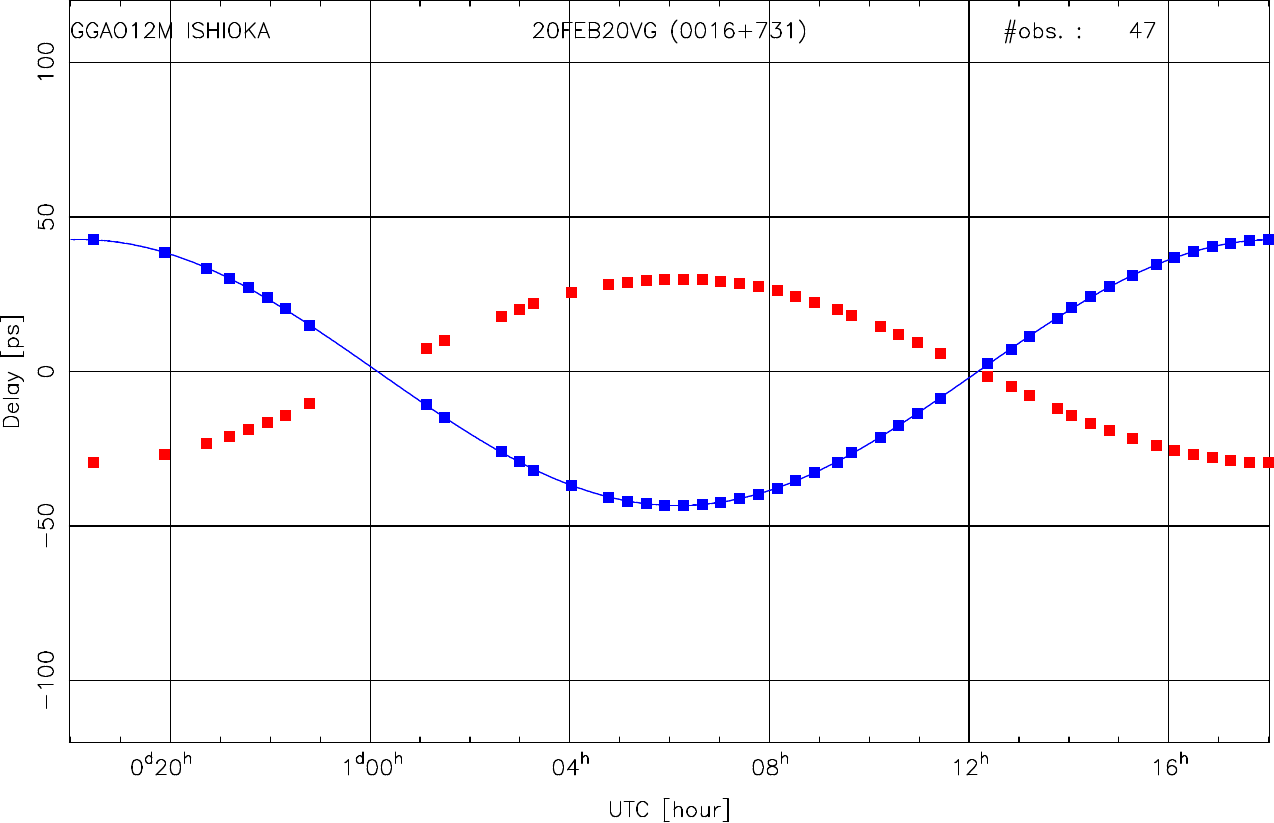}
      \includegraphics[width=0.47\textwidth]{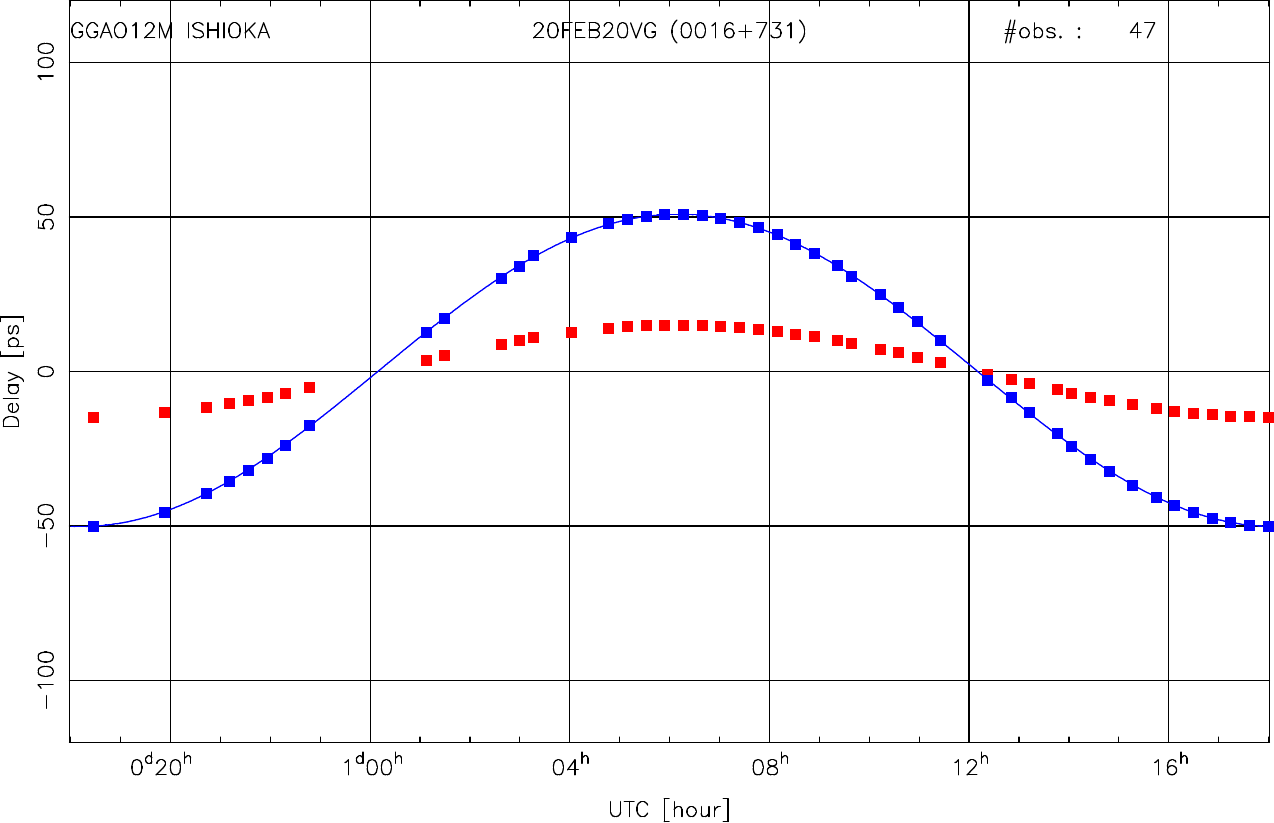}
   \caption{Simulation of broadband group delays (blue dots) of the baseline \texttt{GGAO12M}--\texttt{ISHIOKA} by assuming the position of source 0016$+$731 at band B to be offset by $+$0.2\,mas in declination (left) and the position at band D to be offset by $+$0.1\,mas in declination (right). In these two cases, the reference position is selected to be the positions at the other three bands. The red dots, corresponding to the 47 observations for the source on the baseline in session VO0051, show the delay offsets at band B for the left panel and at band D for the right panel due to the assumed position shifts. The blue dots are the broadband group delays based on the simulation for the 47 observations. The position estimate determined from the simulated broadband delays for the 1308 actual observations of source 0016$+$731 is (0.0, $-0.289$)\,mas for the first case and (0.0, 0.340)\,mas for the second case. The models based on these position estimates are shown as blue curves. }
              \label{src_0016+731}%
    \end{figure*}

Simulation based on the VGOS session VO0051 (20 February 2020) was done to demonstrate the results for group delays. Figure \ref{src_0016+731} shows two cases, where we assume that the position of source 0016$+$731 at band B is offset by 0.2\,mas in declination with respect to that at the other three bands (in the same position and formally selected as $\vec{k}_{0}$) and at band D by 0.1\,mas in declination. By referring to $\vec{k}_{0}$, the position-offset-induced phases at the 32 channels of each individual observation were calculated, from which the broadband observables were fitted using Eq.~\ref{VGOS_structure_effect}. The calculation was done for all the observations of the source 0016+731 in the session one by one. The source positions determined by these simulated broadband group delays for the 1308 observations of the source in the session are (0.000, $-0.289$)\,mas and (0.000, 0.340)\,mas for the two cases, respectively. They can be predicted based on Eq. \ref{group_position}; therefore, the results from the simulation based on actual VGOS observations agree with this equation. In the simulation, the position offset of 0.2\,mas causes delays on a 9000 km baseline with a magnitude of $\sim$30 ps, which is smaller than the phase-delay ambiguity spacings at the four bands and thus does not cause an issue of 2$\pi$ ambiguities in channel phases when doing LSF. This issue could happen in the actual observations if position offsets were larger.

Note that source structure and core shift evolve continuously with frequency, thus leading to additional phase variations within individual bands. These intraband phase variations, compared to the phase variations from band to band (much wider frequency ranges), cause a second-order impact on broadband VGOS observables and are not accounted for in the study.

Two other kinds of simulations were performed the results of which are shown in the Appendix \ref{A1} and \ref{A2}. It is possible that channel frequencies in VGOS observations will be changed to a wider bandwidth and a broader frequency range than the current settings in the future. The results for two other possible sets of channel frequencies are given in Appendix \ref{A1}. The main conclusion from this simulation is that with a broader frequency range the coefficients in Eqs. \ref{group_position} and \ref{phase_position} will be reduced significantly. This also means if the VGOS observing frequencies are changed, for instance, going to 14 GHz, adjusting frequency setups in realtime to avoid RFI at individual stations, or having one band at a station be missing because of RFI or hardware problems, the source position estimates may change substantially. This may have a significant impact on creating and using a VGOS CRF. The results from the simulation with different values of the reference frequency for phase, $\nu_{0}$ in Eq. \ref{VGOS_structure_effect}, are given in Appendix \ref{A2}. The magnitudes of the four coefficients can be significantly reduced for VGOS phase delay positions by increasing the reference frequency for the phase observables in the VGOS data processing.

\section{Discussion}

\subsection{Variations in VGOS source positions}
\begin{figure}[tbhp!]
   \centering
   \includegraphics[width=0.48\textwidth]{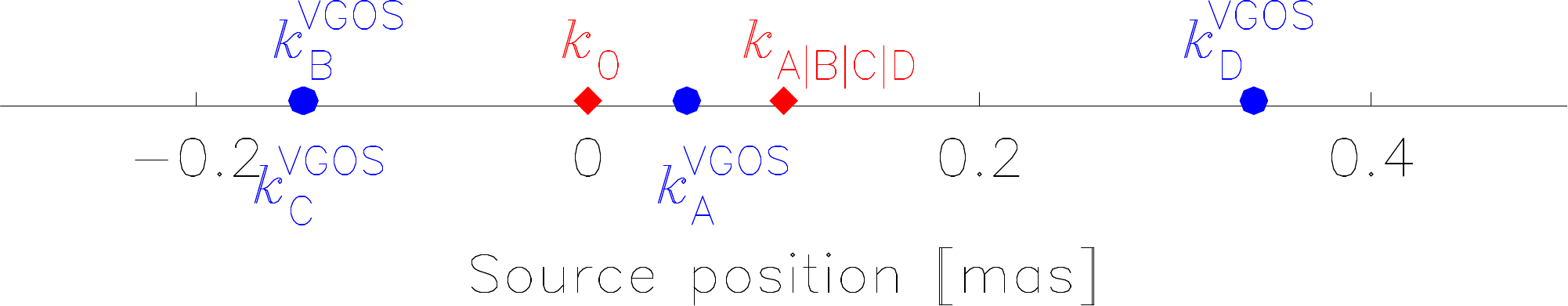}
    
   \caption{One dimensional diagram of the relation between the VGOS group delay positions (blue dots) and the locations of the radio emission at individual bands (red rhombuses). It shows four scenarios: the location of the radio emission at band A, B, C, or D as marked by $\vec{k}_{\scriptsize \mbox{A|B|C|D}}$ is offset by 0.1\,mas with respect to the locations at the other three bands, which are located at the origin marked by $\vec{k}_{0}$. The VGOS group delay positions are shown as blue dots: for example, $\vec{k}^{\scriptsize \mbox{VGOS}}_{\scriptsize \mbox{D}}$ is where the VGOS group delay position is located when the location of the radio emission at band D is offset by 0.1\,mas with respect to the other three bands.}
              \label{vgos_position}%
    \end{figure}

The source positions determined by the VGOS observables are linearly dependent on the locations of the four-band radio emission, as shown in Eq. \ref{group_position} for group delays and in Eq. \ref{phase_position} for phase delays. 
The summations of the four linear coefficients are unity for both cases of group delay and phase delay source positions; however, three of them have absolute values larger than 1, and two have negative values. A major consequence is that if there are position variations due to structure evolution at the three highest bands, they must cause larger variations in the estimates of the VGOS source positions than the actual position variations at individual bands.
Figure \ref{vgos_position} shows a one-dimensional diagram of the VGOS group delay positions in the four scenarios of position offsets between various bands. In three of these four simplest scenarios, the VGOS group delay positions are located far away from the area of the actual radio emission --- $\vec{k}_{0}$ to $\vec{k}_{\scriptsize \mbox{A|B|C|D}}$. This will add complexities to the understanding of the radio and optical position differences in the future. 

Because the core of source 0016+731 at 3.3\,GHz as modeled by difmap is shifted towards east by about 0.14\,mas, the corresponding impact on the VGOS source position can thus be calculated according to Eqs. \ref{group_position} and \ref{phase_position}; it is 0.07\,mas towards east for group delays and 0.12\,mas towards west for phase delays. These impacts may prevent the VGOS from achieving its goal of 1\,mm station position accuracy, and we must emphasize these are the impacts by assuming that the effects of source structure are modeled in the geodetic data analysis. When these effects are not modeled, the impacts are expected to be more complicated and more variable.    

Based on the routine geodetic data analysis of the session VO0051 by using \emph{$\nu$Solve}\footnote{\url{https://sourceforge.net/projects/nusolve/}}, 25 of the 63 sources with more than three observables available have position adjustments with respect to their S/X positions with magnitudes larger than three times the respective uncertainties estimated from the solution. The median angular separation of the positions from VGOS and S/X observations for these 25 sources is 0.387~mas, which has an uncertainty of 0.056\,mas. It is, therefore, a common strategy in analyzing VGOS observations to estimate source positions, which is usually unnecessary in the routine daily solutions of the legacy S/X observations (Sergei Bolotin, private communication).

As demonstrated based on multi-frequency observations for four close CRF sources by \citet{2011AJ....141...91F}, the ICRF position of a source can be dominated by a jet component displaced from the radio core by $\sim$0.5~mas and moving with a velocity of 0.2~mas/yr. This offset and the change will be amplified significantly in VGOS position estimates, leading to much larger position variations in a VGOS CRF than in the previous ICRFs. However, it should be noted that this study is for four ICRF sources only.

%

\subsection{Aligning the images at the four bands}

The simulation so far does not explicitly model the effects of relative source
structure --- the apparent two dimensional distribution of emission on
the sky as a function of frequency, as opposed to an ideal point-source per frequency. Relative source structure affects the phases of
individual frequency channels in a frequency-, baseline length-, and
baseline orientation-dependent manner, and these phases in turn affect
the broadband group delay and phase. Given a model of the relative
source structure, such as images of a source at the four VGOS bands,
the frequency channel phases can be corrected to correspond to the
phases of a point source for each band. However, as images derived
using self-calibration \citep[e.g.,][]{1977Natur.269..764W,1981MNRAS.196.1067C,1984ARA&A..22...97P,2017isra.book.....T} and closure-based imaging \citep{2016ApJ...829...11C,2018ApJ...857...23C} 
techniques lose absolute source position information, it is necessary
to properly align the images at the four bands in order to generate
coherent source-structure phase corrections across all four bands.
Otherwise, a misalignment of the images at the four bands
%
%
%
will introduce a change in the position estimate from the VGOS observables with these corrections applied. This has been demonstrated by \citet{https://doi.org/10.1029/2020JB021238}. The exact formulas of the impact of the misalignment are given above as Eq. \ref{group_position} for group delays and Eq. \ref{phase_position} for phase delays. These results are important for deriving VGOS source-structure corrections. 

The third type of VGOS broadband observables, i.e. $\delta{\mbox{TEC}}$, may also help to determine the differences in source positions at the four bands or align the images over frequency, because these observables are affected by source position offsets as well. Its change due to position offsets among the four bands, denoted by $\Delta{\mbox{TEC}}$, is given by 
\begin{equation}
\label{iono}
\Delta{\mbox{TEC}}= -\vec{B}\cdot(-118.7\Delta \vec{k}_{\scriptsize \mbox{A}}+168.3\Delta \vec{k}_{\scriptsize \mbox{B}}+200.7\Delta \vec{k}_{\scriptsize \mbox{C}}-250.3\Delta \vec{k}_{\scriptsize \mbox{D}})/c,
\end{equation}
where $\Delta{\mbox{TEC}}$ is in units of TECU given that the baseline vector is in units of m, the speed of light is in units of m/s, and the position offsets are in units of mas. This equation is derived from the same way as those for broadband group and phase delays in Sect. \ref{sect3}. The summation of the four coefficients in the right-hand side is zero, therefore, $\Delta{\mbox{TEC}}$ is independent of the reference source position selected. In the case where the position offsets are a function of the frequency to the power of $-1$ as discussed in Sect. \ref{coreshift}, $\Delta{\mbox{TEC}}$ is equal to zero. With external information about the ionospheric effects for VGOS observations available, this quantity may help to validate the image alignment in the presence of core shift.

If there is a position offset at only one of the four bands with a magnitude of 0.1\,mas, $\Delta{\mbox{TEC}}$ for the observations of a 6000 km baseline will have a pattern of a sinusoidal wave over 24 hours of GMST time with a magnitude in the range 0.2--0.5\,TECU based on Eq. \ref{iono}, depending on which band the position offset occurs. Through the comparison of $\delta{\mbox{TECs}}$ from VGOS observations and global TEC maps\footnote{See, for instance, \url{http://ftp.aiub.unibe.ch/ionex/draft/ionex11.pdf}} derived from global navigation satellite systems observations, the precision of $\delta{\mbox{TEC}}$ from global TEC maps is probably on the order of 1--2\,TECU (Zubko et al., in preparation). Therefore, position offsets among the four bands larger than 0.4\,mas may be detectable by comparing $\delta{\mbox{TEC}}$ estimates from VGOS and global TEC maps.

\subsection{Differences in VGOS group delay and phase delay source positions}
\label{coreshift}
Consider the scenario where the position change of a source over frequency $\nu$ is a power-law function $\Delta\mbox{X}/\nu$, where $\Delta\mbox{X}$ is a constant position shift, following the studies from \citet{1984ApJ...276...56M} and \citet{1998A&A...330...79L}. It corresponds to the core shift effect with the frequency power-law of $-1$. At the band with a center frequency $\nu_{\scriptsize \mbox{center}}$, which can be calculated from Table \ref{10GHz}, the source position offset then is $\Delta\mbox{X}/\nu_{\scriptsize \mbox{center}}$. Applying the position offsets in this scenario to Eqs. \ref{group_position} and \ref{phase_position} gives that $\Delta\vec{k}^{\scriptsize \mbox{g}}=0$ and $\Delta\vec{k}^{\scriptsize \mbox{p}}=\Delta\mbox{X}/6.0$. The results show that if core shift is a function of the frequency to the power of $-1$, group delays refer to the position of the AGN jet base, which is the position at the finite frequency, but the phase delays refer to the position of the actual radio emission at the reference frequency. This result was originally derived by \citet{porcas2009}. We can see that Eqs. \ref{group_position} and \ref{phase_position} describe the differences in the reference positions between group delays and phase delays for general scenarios. This generalization is necessary because astronomical results such as \citet{2013A&A...557A.105F} and \citet{2019MNRAS.485.1822P} demonstrate that the position dependence on frequency is often not the power $-1$, and in fact the power index can be variable in time. Instead of considering core shift as a function of frequency, we can describe it directly as position offsets at individual bands.

As shown in Eqs. \ref{group_position} and \ref{phase_position}, the position offset at one band leads to position changes in the opposite directions for the broadband group delays and phase delays. It is likely that these two kinds of broadband observables actually refer to different directions with a quite large separation on the sky. If finally the phase observables from VGOS can be made use of in geodetic solutions as recently did for the very short baselines by \citet{2020arXiv201016214V,2021arXiv210302534N,10.1002/essoar.10507984.1}, the source position differences need to be addressed. Based on Eqs. \ref{group_position} and \ref{phase_position}, the difference of source positions between VGOS group delays and phase delays, denoted by $\Delta\vec{k}^{\scriptsize \mbox{g-p}}$, is given by 

\begin{equation}
\label{group_phase}
\Delta\vec{k}^{\scriptsize \mbox{g-p}}= + 1.388\Delta \vec{k}_{\scriptsize \mbox{A}}-3.177\Delta \vec{k}_{\scriptsize \mbox{B}}-3.501\Delta \vec{k}_{\scriptsize \mbox{C}}+5.290\Delta \vec{k}_{\scriptsize \mbox{D}}.
\end{equation}
Since the four coefficients have the summation of zero, $\Delta\vec{k}^{\scriptsize \mbox{g-p}}$ is a quantity independent of the reference position $\vec{k}_{0}$. The difference in source positions between VGOS group delays and phase delays measures an absolute-scalar product as a combination of the four-band position offsets, including core shift. The measurement noise levels of both these two types of VGOS observables are 1--2 ps or even smaller, which allows this position product to be detected at a few $\mu$as level (Xu et al., in preparation). 

%
%

\section{Conclusion}
We have derived the formulas of the source position estimates from VGOS broadband group delays and phase delays as a function of the locations of the radio emission at the VGOS frequency bands. The resolution across the source is very different at different VGOS bands (a factor of three within the current frequency range and nearly a factor of five between 3 and 14 GHz for the future), and what parts of the jet components contribute to the core flux and position may
vary a lot --- this is likely to move the ``core'' position even further down the jet at lower frequencies (and resolution), especially if the additionally included jet components have steeper spectrum. Source position offsets between various VGOS bands are expected to be common. The variations in the VGOS source positions will be significantly larger than the actual changes in the emission locations due to structure evolution if they happen at the three highest frequency bands. Accordingly to Eqs. \ref{SX_position} and \ref{group_position}, we should expect VGOS sources to wander around parallel to the jet directions as a function of time with much larger magnitudes than the S/X positions did because of variations in source structure and core shift being amplified by the VGOS fringe-fitting strategy. The only way to mitigate these frequency-dependent impacts on VGOS source positions is to have source structure and core shift measured \cite[see, e.g.,][]{2008A&A...483..759K,2011AJ....141...91F,2011A&A...532A..38S} and then correct these effects. This is critical in order to make full use of the high quality data from VGOS with a random noise level of $\sim$2\,ps.

If we want to derive source-structure corrections for VGOS observations, aligning the images at the four bands is very essential: a misalignment can introduce a larger offset in the position estimate from the VGOS observations with these corrections applied than the magnitude of the misalignment itself.

\begin{acknowledgements}
We are grateful to the IVS VGOS stations at GGAO (MIT Haystack Observatory and NASA GSFC, USA), Ishioka (Geospatial Information Authority of Japan), Kokee Park (U.S. Naval Observatory and NASA GSFC, USA), McDonald (McDonald Geodetic Observatory and NASA GSFC, USA), Onsala (Onsala Space Observatory, Chalmers University of Technology, Sweden), Westford (MIT Haystack Observatory), Wettzell (Bundesamt f\"{u}r Kartographie und Geod\"{a}sie and Technische Universit\"{a}t M\"{u}nchen, Germany), and Yebes (Instituto Geogr\'{a}fico Nacional, Spain), to the staff at the MPIfR/BKG correlator center and the MIT Haystack Observatory correlator for performing the correlations and the fringe fitting of the data, to the NASA GSFC VLBI group for doing the geodetic solutions, and to the IVS Data Centers at BKG (Leipzig, Germany), Observatoire de Paris (France), and NASA CDDIS (Greenbelt, MD, USA) for the central data holds.\\
We would like to thank Bill Petrachenko for his helpful comments on this work. We used the Astrogeo VLBI FITS image database for our work, and specifically we thank Bo Zhang for providing the image of source 1157$-$215 at 8.7\,GHz. This research has made use of data from the MOJAVE database that is maintained by the MOJAVE team \citep{2018ApJS..234...12L}. The research was supported by the Academy of Finland project No. 315721 and by the German Research
Foundation grants HE5937/2-2 and SCHU1103/7-2.
\end{acknowledgements}

%
%
%
%
%
%
%
%
%
%
%
%
\bibliographystyle{aa} 
\bibliography{gaia_crf} 

\begin{appendix}
\section{Frequency settings for VGOS observations}\label{A1}
The frequencies of the 32 channels in the current VGOS observations are shown in Table \ref{10GHz}.

The IVS proposes to observe with 1\,GHz wide bands and to extend to higher frequencies, as originally planned \citep{2009vlbi.rept....1P}. 

We first performed the simulation for the frequency range 3.0--11.2\,GHz but with 992~MHz wide bands as listed in Table \ref{10_1GHz} (Bill Petrachenko, private communication). The equations equivalent to to Eqs. \ref{group_position} and \ref{phase_position} are given by 
\begin{equation}
\label{group_position_10_1}
\Delta\vec{k}^{\scriptsize \mbox{g}}_{992,11}= + 0.399\Delta \vec{k}_{\scriptsize \mbox{A}}-1.382\Delta \vec{k}_{\scriptsize \mbox{B}}-1.359\Delta \vec{k}_{\scriptsize \mbox{C}}+3.342\Delta \vec{k}_{\scriptsize \mbox{D}},
\end{equation}

and  
\begin{equation}
\label{phase_position_10_1}
\Delta\vec{k}^{\scriptsize \mbox{p}}_{992,11}=-0.796\Delta \vec{k}_{\scriptsize \mbox{A}}+1.7333\Delta \vec{k}_{\scriptsize \mbox{B}}+2.018\Delta \vec{k}_{\scriptsize \mbox{C}}-1.955\Delta \vec{k}_{\scriptsize \mbox{D}}.
\end{equation}

We then performed the simulation for the frequency range 3.0--14.0\,GHz as listed in Table \ref{14GHz} (Bill Petrachenko, private communication). The results are given as follows:
\begin{equation}
\label{group_position_14}
\Delta\vec{k}^{\scriptsize \mbox{g}}_{992,14}= + 0.248\Delta \vec{k}_{\scriptsize \mbox{A}}-1.008\Delta \vec{k}_{\scriptsize \mbox{B}}-0.995\Delta \vec{k}_{\scriptsize \mbox{C}}+2.755\Delta \vec{k}_{\scriptsize \mbox{D}},
\end{equation}

and  
\begin{equation}
\label{phase_position_14}
\Delta\vec{k}^{\scriptsize \mbox{p}}_{992,14}=-0.633\Delta \vec{k}_{\scriptsize \mbox{A}}+1.663\Delta \vec{k}_{\scriptsize \mbox{B}}+2.078\Delta \vec{k}_{\scriptsize \mbox{C}}-2.108\Delta \vec{k}_{\scriptsize \mbox{D}}.
\end{equation}

There is only a marginal improvement in terms of reducing the magnitudes of the coefficients by using a larger 960\,MHz bandwidth for the current VGOS frequency range. However, significant improvement happens by extending the highest frequency to be 14\,GHz, especially for group delay positions; the magnitudes of the four coefficients are decreased by 20\% to 50\% compared to those of the current frequency setting in Table \ref{10GHz}. This means that the impact of any position offset at a single band has less influence on the VGOS group/phase delay position.

\begin{table*}
\caption{Channel frequencies in the range 3.0--10.7\,GHz currently used in VGOS observations (Units: MHz).}\label{10GHz}
\centering
\begin{tabular}{lrrrrrrrr}
\hline
&1&2&3&4&5&6&7&8\\
\hline
Band A&    3032.4& 3064.4& 3096.4& 3224.4& 3320.4& 3384.4& 3448.4& 3480.4\\
Band B&    5272.4& 5304.4& 5336.4& 5464.4& 5560.4& 5624.4& 5688.4& 5720.4\\
Band C&    6392.4& 6424.4& 6456.4& 6584.4& 6680.4& 6744.4& 6808.4& 6840.4\\
Band D&   10232.4& 10264.4& 10296.4& 10424.4& 10520.4& 10584.4& 10648.4& 10680.4\\
\hline
\end{tabular}
\end{table*}

\begin{table*}
\caption{Channel frequencies in the range 3.0--11.2\,GHz with 992\,MHz wide bands (Units: MHz).}
\label{10_1GHz}
\centering
\begin{tabular}{lrrrrrrrr}
\hline
&1&2&3&4&5&6&7&8\\
\hline
Band A &   3000.4&    3032.4&    3128.4&    3288.4&    3576.4&    3768.4&    3896.4&    3960.4\\
Band B &   5240.4&    5272.4&    5368.4&    5528.4&    5816.4&    6008.4&    6136.4&    6200.4\\
Band C &   6360.4&    6392.4&    6488.4&    6648.4&    6936.4&    7128.4&    7256.4&    7320.4\\
Band D &  10200.4&   10232.4&   10328.4&   10488.4&   10776.4&   10968.4&   11096.4&   11160.4\\
\hline
\end{tabular}
\end{table*}

\begin{table*}
\caption{Channel frequencies in the range 3.0--14.0\,GHz (Units: MHz).}\label{14GHz}
\centering
\begin{tabular}{lrrrrrrrr}
\hline
&1&2&3&4&5&6&7&8\\
\hline
Band A&   3000.4 &   3032.4  &  3128.4  &  3288.4 &   3576.4  &  3768.4 &   3896.4 &   3960.4\\
Band B&   5688.4 &   5720.4  &  5816.4  &  5976.4 &   6264.4  &  6456.4 &   6584.4 &   6648.4\\
Band C&   7832.4 &   7864.4  &  7960.4  &  8120.4 &   8408.4  &  8600.4 &   8728.4 &   8792.4\\
Band D&  13016.4 &  13048.4 &  13144.4  & 13304.4 &  13592.4  & 13784.4 &  13912.4 &  13976.4\\
\hline
\end{tabular}
\end{table*}

\section{Comparison with different values of $\nu_{0}$}
\label{A2}
A simulation was performed to compare the results when varying the reference frequency $\nu_{0}$
in Eq. \ref{VGOS_structure_effect}, rather than the reference value of 6.0\,GHz used in the current fringe fitting by \emph{fourfit}. The relation for group delay positions as Eqs. \ref{delay_offset} and \ref{group_position} does not change when varying $\nu_{0}$, as well as does that for the ionospheric observable. However, the coefficients in the relation of VGOS phase delay positions will decrease by increasing $\nu_{0}$. When setting, for example, $\nu_{0}$=8.5\,GHz, the VGOS phase delay position is given by

\begin{equation}
\label{phase_position_8.5}
\Delta\vec{k}^{\scriptsize \mbox{p}}_{\nu_{0}=8.5~\scriptsize \mbox{GHz}}=-0.475\Delta \vec{k}_{\scriptsize \mbox{A}}+0.794\Delta \vec{k}_{\scriptsize \mbox{B}}+1.014\Delta \vec{k}_{\scriptsize \mbox{C}}-0.333\Delta \vec{k}_{\scriptsize \mbox{D}}.
\end{equation}

By increasing $\nu_{0}$ from 6.0\,GHz to 8.5\,GHz, $\tau^{\scriptsize \mbox{p}}$, which is inversely proportional to $\nu_{0}$, will be simply decreased by 30\%. Without taking this expected decrease into account, the magnitudes of the four coefficients are actually reduced by 16\%, 24\%, 30\%, and 52\%, respectively. However, changing the reference frequency away from the central frequency 6.0~GHz will introduce the errors in group delay estimates into phase estimates. The optimum reference frequency for phases will have to compromise these two factors.
\end{appendix}

\end{document}